\documentclass{article}
\usepackage{amsmath}
\usepackage{amsfonts,amssymb}
\usepackage{psfig}

\newtheorem{thm}{Theorem}
\newtheorem{corr}{Corollary}
\newtheorem{lem}{Lemma}
\newtheorem{prop}{Proposition}

\newcommand{\MM}{\,\mbox{\bf M}}
\newcommand{\HM}{\,\mbox{\bf H}}
\newcommand{\UM}{\,\mbox{\bf U}}
\newcommand{\IM}{\,\mbox{\bf I}}
\newcommand{\JM}{\,\mbox{\bf J}}

\newcommand{\U}{\mathbb{U}}
\newcommand{\tr}{\,\mbox{\rm Tr}}
\newcommand{\Lp}{{\bf L_+}}
\newcommand{\Lm}{{\bf L_-}}
\newcommand{\Lpm}{{\bf L_\pm}}
\newcommand{\tphi}{{\tilde\phi}}
\newcommand{\sgn}{\operatorname{sgn}}
\newcommand{\sn}{\operatorname{sn}}
\newcommand{\cn}{\operatorname{cn}}
\newcommand{\dn}{\operatorname{dn}}
\newcommand{\spec}{\operatorname{spec}}

\psfull

\begin{document}

\bibliographystyle{plain}
\nocite{*}

\centerline{\bf Modulational Instability for Nonlinear Schr\"odinger  
Equations with a Periodic Potential. }
 
\centerline{\bf Jared C. Bronski\footnote{ Department of Mathematics,
University of Illinois Urbana-Champaign, 1409 W. Green St., Urbana, 
IL 61820, email:jared@math.uiuc.edu, FAX: (217) 333-9576}}
\centerline{\bf Zoi Rapti\footnote{ School of Mathematics, Institute 
for Advanced Study, Einstein Drive, Princeton, NJ 08540, email: 
rapti@math.ias.edu}}

\begin{abstract}

We study the stability properties of periodic solutions to the 
Nonlinear Schr\"odinger (NLS) equation 
with a periodic potential.
We exploit the symmetries of the problem, in particular the 
Hamiltonian structure and the $\U(1)$ symmetry. 
We develop a simple sufficient criterion that guarantees 
the existence of a modulational instability spectrum 
along the imaginary axis. In the case of small amplitude 
solutions that bifurcate from the band edges of the linear 
problem this criterion becomes especially simple. 
We find that the small amplitude solutions corresponding to the 
band edges alternate stability, with the first band edge being 
modulationally unstable in the focusing case, the second band edge 
being modulationally unstable in the defocusing case, and so on.
This small amplitude result has a nice physical interpretation in terms
of the effective mass of a particle in the periodic potential. 
We also consider, in somewhat less detail, some sideband 
instabilities in the small amplitude limit. We find that, 
depending on the Krein signature of the collision, these
can be of one of two types. 
Finally we illustrate this with some exact calculations in the case where 
$V(x)$ is an elliptic function, where all of the relevant calculations 
can be done explicitly.

\end{abstract}

\section{Introduction}

In this paper we consider the stability of standing wave solutions to 
the NLS equation with an external potential:
\begin{equation}
i\psi_t = -\frac{1}{2} \psi_{xx} \pm \epsilon |\psi|^2 \psi + V(x) \psi,
\end{equation}
where the potential $V(x)$ has period $1$. 
We assume that we have a standing wave solution $\psi(x,t,\epsilon) = 
\exp(-i \omega(\epsilon) t ) \phi(x,\epsilon)$, where $\phi(x,\epsilon)$
is real valued and either periodic ($\phi(x+1,\epsilon) =  \phi(x,\epsilon)$) 
or antiperiodic 
($\phi(x+1,\epsilon) =  -\phi(x,\epsilon)$). For certain perturbative 
results we will also assume that the standing wave 
bifurcates continuously from the Floquet-Bloch eigenfunctions in 
the usual way. In other 
words we assume that $\phi(x,0)$ is an eigenfunction of the periodic 
Schr\"odinger operator 
\begin{equation}
\mu \phi = \omega(0) \phi =  -\frac{1}{2} \phi_{xx}  + V(x) \phi
\end{equation}
corresponding to a band edge, and $\phi(x,\epsilon)\in C^2\left([0,1]\times
(-\delta,\delta)\right)$ 

We find that the combination of the Hamiltonian structure 
and the $\U(1)$ symmetry dramatically simplifies the 
structure and possible bifurcations of the spectrum of the 
linearized operator. We find a simple condition on the ${\bf L}_+$ 
operator which guarantees the existence of a modulational 
instability. 
In the case of weakly nonlinear standing waves, 
when $\epsilon$ is small, we apply a perturbation argument to show that 
the lower band edges are modulationally unstable in the focusing case, 
while the upper band edges are modulationally unstable in the defocusing 
case. 

We also study, in somewhat less detail, some bifurcations which occur for 
non-zero values of $\mu$ when a certain eigenvalue degeneracy condition 
is met. We find that this bifurcation generically leads to the emergence of 
complex eigenvalues, and thus instability. Depending on the Krein signature 
of the unperturbed eigenvalues this may or may not lead to the opening of 
a gap along the real axis. 

\section{Fundamentals of Systems with Symplectic Structure and 
Periodic Coefficients}

In this section we review some results of Floquet theory for the special 
case in which the equations admit a Hamiltonian formulation. For 
more details see the text of Yakubovich and Starzhinskii\cite{YS}. 

Throughout this paper we will apply the following notation. 
${\bf U}(x)$ will denote the solution operator of the periodic 
ODE governing stability. The matrix $\MM$ will denote the monodromy, or 
period, map ${\bf U}(1)$. We will have occasion to consider the 
spectral properties of the second order operators ${\bf L_\pm}$ 
individually. The matrix 
${\bf m}$ will denote the mondromy matrix of the second order problem
associated to ${\bf L_\pm}$. 
For each of these quantities the dependence on the parameters
$\mu,\epsilon$ will be generally be suppressed unless it is 
necessary for clarity. Similarly $K_\pm(\mu)$ will denote the 
Floquet discriminants of the full stability problem, while 
$k(\mu)$ will denote the Floquet discriminant of the second order problem
associated with ${\bf L_\pm}$.   

\subsection{ The Lyapunov-Poincare theorem}

To begin we assume a Floquet problem of the following form:
\begin{equation}
\UM_x = \JM \HM(x) \UM ~~~~ \UM(0) = \IM_{2N\times 2N}~~~~\HM(x+1) = 
\HM(x)~~~\HM^t = \HM
\end{equation}
where, for simplicity, we have set the period to one. Here 
$\IM_{2N\times 2N}$ is the $2N\times 2N$ identity matrix and $\JM$ 
is a skew-symmetric matrix. We will assume that $\JM^2 = -\IM$ 
since the above problem with a non-singular $\JM$ can always 
be mapped to one of that form. The monodromy 
matrix $\MM$ is defined to be the period map $\MM=\UM(1)$. It is easy to 
see that $\UM(x)$, and thus $\MM$, satisfies the relation 
\begin{equation}
\UM^t \JM \UM = \JM,
\end{equation}
so that  $\UM(x)$ is a symplectic\footnote{Sometimes called $\JM$-unitary} 
matrix. 
From this it follows that $\det(\UM)=1$, and more generally that the 
characteristic polynomial $P(\lambda) = \det(\UM-\lambda \IM)$ satisfies
\begin{eqnarray}
\det(\UM - \lambda \IM) &=& \det(\UM^t - \lambda \IM) =\det(\JM (\UM^t - 
\lambda \IM) \JM^t) \\
&=& \det(\UM^{-1} - \lambda \IM)  = \lambda^{2N} \det(\UM^{-1}) \det(\UM - 
\lambda^{-1} \IM) \\
&=& \lambda^{2N} P(\lambda^{-1})
\end{eqnarray}
thus the polynomial is palindromic - if $P(\lambda) = 
\sum_0^{2N} a_j \lambda^j$ then $a_{2N-j} = a_j$. This result is known as the 
Lyapunov-Poincare theorem \cite{YS}. 

This symmetry implies that the problem of finding the roots of the 
$2N^{\rm th}$ degree polynomial 
$P(\lambda)$ can  be reduced to finding the roots of an $N^{\rm th}$ degree 
polynomial $\tilde P(z)$ by means of the transformation $z = \lambda + 
\lambda^{-1}$. 
For the case of the stability problem for standing wave solutions to 
the nonlinear Schrodinger equation the monodromy matrix is 
$4 \times 4\,\, (N=2)$ 
In the case the characteristic polynomial of the monodromy matrix 
takes the form
\begin{equation}
P(\lambda) = 1 + a \lambda + b \lambda^2 + a \lambda^3 + \lambda^4 = 0.
\end{equation}
Under the conformal map the roots of this polynomial are 
mapped to the roots of the second degree polynomial 
\begin{equation}
\tilde P(z) = z^2 + a z + (b-2) = 0
\end{equation}
with $\lambda = \frac{z \pm \sqrt{z^2 - 4}}{2}$, and the 
characteristic polynomial 
admits an explicit factorization into two second degree polynomials,
\begin{equation}
P(\lambda) = (1 - K_+ \lambda + \lambda^2)(1 - K_-\lambda + \lambda^2)
\label{eqn:fact1}
\end{equation}
where the $K_\pm$ are the following algebraic functions of $a,b$:
\begin{equation}
K_\pm = \frac{-a \pm\sqrt{a^2 - 4 b + 8}}{2}.
\label{eqn:fact2}
\end{equation}
The functions $K_\pm$ will be refered to as the Floquet discriminants. 
Since the conformal mapping $z= \lambda + \frac{1}{\lambda}$ takes the unit 
circle to the real interval $[-2,2]$ it follows that  
the monodromy matrix has two eigenvalues (counted by algebraic multiplicity)
on the unit circle if $K_+$ lies in the real interval $[-2,2]$, and two 
more if $K_- \in [-2,2]$. 

It is convenient to express the 
$K_\pm$ in terms of the invariants of $\MM$. Since $a = - \tr(\MM), b =- 
\frac{1}{2}\tr(\MM^2)+\frac{1}{2} \tr(\MM)^2$ the Floquet discriminants 
are given by 
\begin{equation}
K_\pm = \frac{\tr(\MM)\pm \sqrt{- (\tr(\MM))^2 + 2 \tr(\MM^2) + 8}}{2}
\label{eqn:fact3}
\end{equation}
This leads to our first observation 
\begin{lem}
The possible (algebraic) multiplicities of the eigenvalues of a
$4 \times 4$ monodromy matrix are $(1,1,1,1)~$, $(1,1,2)$, $(2,2)$ and $4$. 
The conditions on 
the monodromy matrix $\MM$ which produce eigenvalues of higher 
multiplicity are as follows:\\
\begin{tabular}{||l|l|l||}\hline\hline
Condition on $K_\pm$ & Condition on $\MM$ &  Root Multiplicities \\\hline
$K_+ = \pm 2$ or $K_- = \pm 2$ & $\tr(\MM^2) = \left(\tr(\MM)\pm 2)\right)^2$
& (1,1,2)\\
Simple Band Edge & & \\ \hline
$K_+ = \pm 2$ and $K_- = \mp 2$ &$\tr(M)=0$ & (2,2) \\
Repeated Band Edge &$\tr(M^2)=4$  & \\ \hline
$K_+=K_- \neq \pm 2$ &  $2\tr(M^2) = (\tr(M))^2 -8 $& (2,2) \\
 `Accidental' Degeneracy &$\tr(M)\neq \pm 4$ & \\ \hline
$K_+=K_- = \pm 2$ & $\tr(M)=\pm 4$& 4 \\
Double Band Edge &$\tr(M^2) = 4$ &  \\ 
\hline\hline
\end{tabular}
\end{lem}

{\it Proof:} A straightforward calculation from the 
explicit factorization given in \eqref{eqn:fact1}. 

It is worthwhile to consider the codimensions of the above possibilities. 
Since we have a one parameter family of mondromy matrices, 
parameterized by the spectral parameter $\mu$, we should generically 
expect to see possibilities 1 and 3, which require only one 
condition, but not possibilities 2 and 4, which require two independent
conditions be satisfied. We shall see that possibility 4 {\em always}
happens at $\mu=0$ for small amplitude waves, and is forced by symmetries.
It is this degeneracy which leads to the modulational instability. It is this 
situation that we will consider in most detail. We will also consider 
possibility 3, in somewhat less detail. We will not consider possibility 
2, since it is non-generic, nor possibility 1, since it is not very 
interesting - one expects that the band edges will typically move under 
perturbation.

\subsection{Krein Signature}

There is an extensive theory of the structural stability of 
symplectic matrices due to Krein and collaborators. We shall 
give only a brief summary of this theory here. It is clear that, 
if a symplectic matrix $\MM$ has $k$ distinct eigenvalues on the 
unit circle it follows from continuity that
 a nearby symplectic matrix $\tilde \MM$ 
must also have  $k$ distinct eigenvalues on the unit circle. 
Thus eigenvalues of symplectic matrices can only leave the 
unit circle via collisions. What is less clear is that only 
certain collisions can lead to pairs of eigenvalues 
leaving the unit circle. The quantity which distinguishes
such ``dangerous'' collisions is  
that of the Krein signature. The Krein sign $\eta$ of an 
eigenvector $\vec v$ of a symplectic matrix $\MM$ is defined to 
be 
\begin{equation}
\eta = \sgn (<\!\!\vec v \JM \vec v\!\!>),
\end{equation}
while the signature $(p, q)$ of an $r-$dimensional eigenspace
is defined as follows: $p$ (resp. $q$) is the number of linearly 
independent eigenvectors with positive (resp. negative) Krein sign.

The fundamental stability result for symplectic matrices says that 
a symplectic matrix is structurally stable (i.e. the number of 
eigenvalues on the unit circle does not change under perturbation
\footnote{Here perturbations are always assumed to preserve the 
symplectic nature of the matrix.})
if all eigenspaces corresponding to eigenvalues on the unit circle 
have a definite Krein signature. Further this result is tight: 
if an eigenspace is of indefinite Krein signature then a generic 
perturbation will cause roots to move off of the unit circle. 
The first result is known as Krein's theorem, while the second is known
as the Krein-Gelfand-Lidskii strong stability theorem. For precise 
statements, as well as proof, see Yakubovich and Starzhinskii\cite{YS}.

\subsection{Hill's Equation}

In this section we state some properties of the Hill's 
equation
\begin{equation}
{\bf H} \psi = -\psi_{xx} + Q(x) \psi = \mu \psi
\label{eqn:floq}
\end{equation}
that will be useful in the sequel. We will also define a quantity 
that will be useful in the analysis to follow. 

In the remainder of this section ${\bf m}$ is the 
($2\times2$) monodromy matrix 
associated to the Hill's equation (\ref{eqn:floq}):
\begin{equation}
{\bf m} = \left( \begin{array}{cc} \psi_1(1,\mu) & \psi_2(1,\mu) \\
\psi_1^\prime(1,\mu) & \psi_2^\prime(1,\mu) \end{array}  \right),
\end{equation}
with $k(\mu) = \tr({\bf m})$, and ${\bf j}$ is the standard skew-symmetric 
form 
\begin{equation}
{\bf j} = \left( \begin{array}{cc} 0 & -1 \\ 1 & 0 \end{array} \right). 
\end{equation}

We begin with some well-known facts. 
The spectrum of the above eigenvalue problem consists of the 
union of the set of intervals $\spec(H)=[\mu_0,{\mu_1}^\prime] \cup 
[{\mu_2}^\prime,\mu_1] \cup [\mu_2,{\mu_3}^\prime] \ldots,$ where the $\mu_i$ 
are the roots of 
$k(\mu) = 2$ and the ${\mu_i}^\prime$ are the roots of $k(\mu)=-2$.   
The points $\lbrace \mu_i\rbrace$ are the periodic eigenvalues, while
the points $\lbrace {\mu_i}^\prime\rbrace$ are the anti-periodic eigenvalues. 
Together these comprise the band edges. It is always true that 
$\mu_{2i} < {\mu_{2i+1}}^\prime$ and $\mu_{2i}^\prime
 < {\mu_{2i-1}}$, so the bands are always nontrivial. It can happen that 
$\mu_{2 i -1}^\prime =  \mu_{2 i}^\prime$ or $\mu_{2 i -1} =  \mu_{2 i}$ 
corresponding to a closed gap.  This is commonly referred to as a double 
point. 
In the interior of each band there are two quasi-periodic solutions, 
while at the band edge there is one 
periodic or antiperiodic solution and one solution which grows linearly 
unless the band edge is a double point, in which case there are two 
periodic or antiperiodic solutions. The derivative of the Floquet 
discriminant is 
non-zero in the bands, has exactly one zero in each gap, and is nonzero on 
the band edges unless the band edge is a double point, in 
which case the derivative of Floquet discriminant has a simple 
zero.  Proofs of these 
facts can be found in Magnus and Winkler\cite{MagWin}. 

In the first lemma we introduce a quantity which will play an important 
role in the perturbation analysis. The sign of this quantity will be 
important  in determining the modulational stability of the standing wave. 

\begin{lem}
Define the quantity $\sigma(\mu)$ as follows:
\begin{equation}
\sigma(\mu) = \tr\left({\bf j m}(\mu)\right)
\end{equation}
Then the following hold
\begin{itemize}
\item $\sigma(\mu)$ is non-zero on the interior of each band.
\item $\sigma(\mu)$ vanishes at a band edge iff the band edge is a double 
point.
\item $\sgn(\sigma(\mu))=-\sgn(k^\prime(\mu))=-\sgn(\tr(\frac{\partial{\bf m}}
{\partial \mu}))$ if $\mu \in \spec({\bf H}).$
\end{itemize}

{\it Proof:} 
On the interior of a band it holds  $k^2(\mu)-4 < 0$. If we 
set $\eta_1=\psi_1(1,\mu)$, $\eta_2=\psi_2(1,\mu)$, 
$\eta_1^{\prime}=\psi_1^\prime(1,\mu)$, and 
$\eta_2^\prime=\psi_2^\prime(1,\mu)$, then  
\begin{equation}
k^2(\mu)-4=(\eta_1-\eta_2^\prime)^2+4 \eta_1^\prime \eta_2, 
\label{disc}
\end{equation}
which implies that
$\eta_1^\prime \eta_2 \le 0$. By its definition, 
$\sigma(\mu)=-\eta_1^\prime+\eta_2=-\sgn(\eta_1^\prime) (|\eta_1^\prime|+
|\eta_2|)$.
But (\ref{disc}) implies that $\eta_1^\prime \ne 0$ and $\eta_2 \ne 0$.
Thus it follows that $\sigma(\mu)$ is nonzero in the interior of each band.

Let a band edge be a double point. It follows that $k^2(\mu)-4= 0$ and 
moreover $k^\prime(\mu)=0$. Following \cite{MagWin} one can show that
\begin{equation}
k^\prime(\mu)=\sgn(\eta_1^\prime)(|\eta_1^\prime| \int \psi_2^2+|\eta_2| 
\int \psi_1^2
\pm 2 \sqrt {|\eta_1^\prime|}\sqrt{|\eta_2|} \int \psi_1 \psi_2), 
\end{equation}

where $\psi_1$, $\psi_2$ are the two linearly independent solutions of 
\begin{equation}
\psi_{xx} + (\mu -Q(x)) \psi =0,
\end{equation}
that satisfy the initial conditions $\psi_1(0,\mu)=1$, 
$\psi_1^\prime(0,\mu)=0$,
$\psi_2(0,\mu)=0$, and $\psi_2^\prime(0,\mu)=1$.
We used again (\ref{disc}) and, in particular, the inequality 
$\eta_1^\prime \eta_2 \le 0$. 
By Cauchy-Schwartz it 
follows that $\sgn(\eta_1^\prime)=0$, thus $\eta_1^\prime=0$. But then, also 
$\eta_1-\eta_2^\prime=0$, which in turn forces $\eta_2=0$.

The other implication is proved as follows: $\sigma(\mu)=0$ implies
$0=k^2(\mu)-4=(\eta_1-\eta_2^\prime)^2+4 \eta_2^2$, so $\eta_2=\eta_1^\prime=0$
and $\eta_1-\eta_2^\prime=0$. But then clearly $k^\prime(\mu)=0$.

Finally, again in \cite{MagWin} it was shown that if $\eta_1^\prime \ne 0$ then
$k^\prime(\mu)$ and $\eta_1^\prime$ have the same sign. The result follows, 
since from the first part of this lemma, $\sigma(\mu)$ and $\eta_1^\prime$ 
have opposite signs.
 
\end{lem}

This quantity is related to the Krein sign, and actually agrees with the 
Krein sign of the eigenvalue of the monodromy matrix in the upper half plane. 
However the main utility 
of this quantity is that it allows one to compute the sign of the 
off-diagonal piece of the Jordan normal form at a band edge, as 
in the lemma below.

\begin{lem}
At a band edge the monodromy matrix ${\bf m}$ has the following Jordan normal 
form:
\begin{equation}
{\bf m} = {\bf o}^t \left( \begin{array}{cc} \pm 1 & \sigma(\mu) \\ 0 & \pm 1  
\end{array}\right) {\bf o}
\end{equation}
where ${\bf o}$ is a proper orthogonal matrix: ${\bf oo^t} = \IM,~~~ 
\det({\bf o})=+1$

{\it Proof:} 
It is known that at a band edge that is not a double point
$\pm 1$ is an eigenvalue of algebraic multiplicity two, and 
geometric multiplicity $1$. The Jordan normal form implies 
that 
\begin{equation}
{\bf m} = {\bf o}^t \left( \begin{array}{cc} \pm 1 & K \\ 0 & \pm 1  
\end{array}\right) {\bf o} = {\bf o}^t \tilde {\bf m} {\bf o} 
\end{equation}
where ${\bf o}$ is orthogonal and can be chosen to have determinant $+1$. 
It remains to be checked that $K = \sigma(\mu)$. One easily observes
that ${\bf j}$ is a rotation and thus commutes
with ${\bf o}$. This implies that $\sigma = \tr({\bf j m}) = 
\tr({\bf o j o^t} \tilde {\bf m}) =  K$. If ${\bf o}$ is chosen to 
be on the other connected component of the orthogonal group 
${\bf j}$ and ${\bf o}$ anti-commute, and the sign of the off-diagonal 
element is reversed.
 
\end{lem}

The value of $\sigma$ at a band edge, in particular the sign, is important in 
determining the stability of small amplitude standing wave solutions. 
Note that the Krein signature of an eigenvalue 
of the monodromy matrix in the upper half-plane is equal to the sign 
of $\sigma$, and thus has the opposite sign from the derivative of the Floquet 
discriminant.

\section{Modulational Instability of Standing Waves}

\subsection{General Results}

We assume that there exists a $C^2$ family of standing wave solutions 
$\psi = e^{-i\omega(\epsilon)t}\phi_{stand}(x,\epsilon)$ 
to the NLS equation with a periodic potential
\begin{equation}
i\psi_t = -\frac{1}{2}\psi_{xx} + \epsilon |\psi|^2 \psi + V(x) 
\psi ~~~~~V(x+1)=V(x),
\end{equation}
which is either periodic $\phi(x+1)=\phi(x)$ or anti-periodic $\phi(x+1)
=-\phi(x)$.  
The eigenvalue problem governing the linearized stability is given by
\begin{eqnarray*}
\Lp p &=& -\mu q \\
\Lm q &=& \mu p
\end{eqnarray*}

where the operators $\Lpm$ are given by
\begin{eqnarray*}
\Lm &=& -\frac{1}{2} \partial_{xx} + V(x) + \epsilon |\phi_{stand}|^2(x,
\epsilon) - 
\omega(\epsilon) \\
\Lp &=& -\frac{1}{2} \partial_{xx} + V(x) + 3 \epsilon |\phi_{stand}|^2(x,
\epsilon) - \omega(\epsilon) 
\end{eqnarray*}
The above eigenvalue problem has a Hamiltonian formulation for 
real $\mu$. The Lyapunov-Poincare theorem of the previous section implies 
that the characteristic polynomial of the monodromy matrix is palindromic,
$P(\lambda) = \lambda^4 P(\lambda^{-1})$, for $\mu$ on the real axis. 
Also note that for 
{\em arbitrary} values of $\epsilon$ the spectrum is symmetric about the 
real and imaginary axes, since the eigenvalue problem is invariant under 
the transformations $\mu \rightarrow -\mu, p \rightarrow -p, q \rightarrow q$
and  $\mu \rightarrow \bar\mu, p \rightarrow \bar p, q \rightarrow \bar q$.

We begin by noting a few simple properties of this eigenvalue problem, and 
the associated monodromy matrix:
 
\begin{prop}
The Floquet stability problem for an NLS standing wave has the following 
properties:
\begin{itemize}
\item $\mu=0$ is a periodic (anti-periodic) eigenvalue. 
\item $\MM(\mu)$ is an entire function of $\mu$, of order $\frac{1}{2}$.
\item The monodromy matrix $M$ is symplectic for all $\mu \in {\bf C}$.
\item The Floquet discriminants $k_\pm(\mu)$ are analytic functions 
of $\mu$ away from the branch points where $2 \tr(\MM^2) - \tr(\MM)^2 = 8$.
\item At $\mu=0$ the characteristic polynomial of the monodromy matrix 
has the following special form:
\begin{equation}
P(\lambda)|_{\mu = 0} = 1 - \tr(\MM) \lambda + (2 \tr(\MM) - 2) \lambda^2 
- \tr(\MM) \lambda^3 + \lambda^4
\end{equation}

\end{itemize}
\end{prop}

{\it Proof:}

That $\mu = 0$ is always an eigenvalue follows from Noether's theorem and 
the phase invariance of NLS. The corresponding eigenvector is  
$p = \phi_{stand}(x), q=0$. The fact that $\MM$ is entire follows from 
standard arguments - see for example the text of Sibuya\cite{Sib}. 
Note that $2 \tr(\MM^2) - \tr(\MM)^2 = 8$ is also an entire function of 
fractional order, and thus must have a countable number of zeros. 
The fact that $\MM$ is symplectic for 
{\it real} $\mu$ follows from the results cited in the previous section.
To see that this in fact holds for all $\mu \in {\bf C}$ simply note 
that $\JM - \MM^t \JM \MM$ is an entire function that is zero on the 
real axis, and thus must be identically zero. 
A simple division shows that 
the polynomial $\lambda - 1$ divides the polynomial
$\lambda^4 + a \lambda^3 + b \lambda^2 + a \lambda + 1$ if and only if
$b+2a+2=0$, which proves the last part. Note that this same 
calculation shows that if $\lambda=1$ is a root it is necessarily 
of multiplicity $2$ or $4$.

\begin{lem} If $0$ is not a periodic eigenvalue of the 
${\bf L}_+$ operator the Floquet discriminants $K_\pm(\mu)$ 
are analytic in a neighborhood of $\mu=0$.
\end{lem} 
{\it Proof:} For $\mu=0$ the stability problem decouples, and the 
monodromy matrix takes the block diagonal form
 \begin{equation}
\MM = \left(\begin{array}{cc}{\bf m_-} & 0 \\ 0 & {\bf m_+} \end{array}\right),
\end{equation}
where ${\bf m_\pm}$ are the monodromy matrices associated with 
${\bf L_\pm}$. 
It follows from the previous lemma that $\lambda =1$ is an 
eigenvalue of ${\bf m_-}$ with multiplicity $2$. A short calculation 
using the fact that $2\times2$ matrices satisfy $\tr({\bf m}^2) = 
\tr({\bf m})^2 - 2 \det({\bf m})$ shows that $2 \tr(\MM^2) - \tr(\MM)^2 + 8 = 
(\tr({\bf m_-})-2)^2$. If  $0$ is not a periodic eigenvalue of ${\bf L}_+$ 
then $2 \tr(\MM^2) - \tr(\MM)^2 + 8\neq 0$ and $K_\pm$ are analytic in 
a neighborhood of $\mu=0$.

\begin{thm} For {\it general} $\epsilon$ a sufficient condition 
for the existence of a modulational instability spectrum is 
that $\mu=0$ is in the interior of a band of the ${\bf L_+}$ 
operator. 
\end{thm}
{\it Proof:} 
From the previous lemma if $\mu=0$ is in the interior of a band of the 
${\bf L_+}$ then the Floquet discriminants $K_\pm(\mu)$ are analytic 
in a neighborhood of the origin. From the fact that the coefficients 
of the characteristic polynomial are invariant under the transformation
$\mu \rightarrow -\mu$ it follows that  $K_\pm(\mu)$ are even, and 
thus are real on an segment of the imaginary axis containing the origin. 
Since $K_-(0) \in (-2,2)$ it follows that $K_-(\mu)$ is real and $ \in (-2,2)$
on some segment of the imaginary axis containing the origin, and thus there 
is a modulational instability.

{\bf Remark:} The same argument guarantees the existence of a modulational 
instability if either $K(0)=+2$ and $K''(0)\ge 0$ or $K(0)=-2$ and 
$K''(0)\le 0$. The calculation of the sign of the second derivative 
appears to require a difficult second order perturbation calculation.

\subsection{Perturbative Results for weak nonlinearity}

In this section we study the Floquet spectrum for small amplitude 
standing waves of the nonlinear Schrodinger equation with a periodic 
potential, with a particular emphasis on the behavior near $\mu=0$. 
We shall see that the ${\bf U}(1)$ phase invariance of the 
NLS forces a four-fold degeneracy of the eigenvalues of the 
unperturbed monodromy matrix at $\mu=0$. Under perturbation this 
degeneracy can lead to the birth of a spine of continuous spectrum 
lying along the imaginary axis. Whether or not such a spine is born 
is determined by the relative sign of the nonlinearity and the 
quantity $\sigma$ defined in the previous section, and the length of 
the spine is determined by the magnitude of $\sigma$.

When $\epsilon=0$ the operators $\Lpm$ are equal and are given by
\begin{equation}
\Lm(0) = \Lp(0) =  -\frac{1}{2} \partial_{xx} + V(x) - \omega(0)
\end{equation}
and the resulting eigenvalue problem is self-adjoint. It is easy to 
see that in this case the spectrum of the stability problem 
is given by $\spec(\Lm) \cup \spec(-\Lm)$. In this case it is 
also straightforward to calculate that the monodromy matrix $\MM$ of the
full stability problem takes the block diagonal form  
\begin{equation}
\MM = \left( \begin{array}{cc}
{\bf m}(\mu) & 0 \\ 0 & {\bf m}(-\mu)\end{array}\right)
\end{equation}
where ${\bf m}(\pm \mu)$ is the monodromy matrix for the  problem 
\begin{equation}
\Lm(0)\psi = \pm \mu \psi. 
\end{equation}
From this block diagonal form it follows that for $\epsilon=0$ the invariants 
of the full monodromy matrix can be expressed in terms of the Floquet
discriminants $k(\pm \mu)$ of the second order problem via
\begin{eqnarray*}
\tr(\MM(\mu)) &=& \tr\left( {\bf m}(\mu)\right) +\tr \left({\bf m}(-\mu)\right)
 = k(\mu) + k(-\mu) \\
\tr(\MM^2(\mu)) &=& \tr\left( {\bf m}^2(\mu)\right) +\tr \left({\bf m}^2(-\mu)
\right) = 
k^2(\mu) + k^2(-\mu) - 4. 
\end{eqnarray*}
Here we have used the fact that $2\times2$ matrices satisfy the 
identity $\tr( {\bf m}^2) = \tr({\bf m})^2 - 2 \det({\bf m})$. From 
this it follows that the Floquet discriminants  
$K_\pm(\mu)$ for the full problem can be written  
in terms of the Floquet discriminant of $\Lm$ via 
\begin{equation}
K_\pm(\mu) = \frac{k(\mu) + k(-\mu) \pm \sqrt{(k(\mu) - k(-\mu))^2}}{2}.
\end{equation}
Obviously this could be simplified to $K_\pm(\mu) = k(\pm\mu)$ however 
we do not do this, since it obscures the degeneracy at 
the points where $k(\mu)=k(-\mu).$
 
The $\U(1)$ symmetry of the NLS equation implies that 
$\mu=0$ is a band-edge eigenvalue of $\Lm$.
Thus for $\epsilon=0,\mu=0$ the monodromy matrix has has a single 
eigenvalue $\pm1$ of multiplicity four. This corresponds to the 
last entry in the table in figure 1. This seemingly non-generic 
bifurcation is forced by the phase-invariance symmetry along with 
the symplectic nature of the monodromy matrix, and  can give rise to the 
modulational instability at non-zero wave amplitudes.

In the next lemma we present a normal form calculation for the 
Floquet discriminants in a neighborhood of 
$\mu = 0, \epsilon =0$. The calculation is particularly simple
due to some additional symmetries, which dramatically reduce the 
number of coefficients which need to be computed. 
We present the 
calculation for periodic band edges: the calculation for the 
anti-periodic bands edges is identical except for some sign changes. 
In this latter case we merely state the final result. 

\begin{lem}
The Floquet discriminants of the monodromy matrix have the 
following normal form in a neighborhood of $\epsilon=0,\mu=0$:
\begin{eqnarray*}
&K_\pm(\mu) = 2 + \frac{k''(0)}{2} \mu^2 + 2 \epsilon \sigma 
<\!\!\phi_1^4\!\!> + E_1
\pm \sqrt{\left(k'(0)\mu\right)^2  + \left(2 \sigma
\epsilon\!<\!\!\phi_1^4\!\!>\right)^2 + E_2}& \\
&E_1 = o(\epsilon,\mu^2)& \\
&E_2 = o(\epsilon^2,\mu^2)&  
\end{eqnarray*}
where $k(\mu)$ is the Floquet discriminant for the ${\bf L_-}(0)$
operator.
\end{lem}

{\em Proof:}
This perturbation calculation is somewhat tedious but can be made 
simpler by the use of some of the previously derived identities. 
From the last part of 
Lemma 4 we can express the $\epsilon$ derivatives of the 
coefficients of the characteristic polynomial at the origin in terms of 
$\tr(\MM|{\mu=0}).$ To compute the $\mu$ derivatives of the coefficients 
of the characteristic polynomial at the origin we use the fact that 
$a(\mu,0) = -(k(\mu)+k(-\mu)), b(\mu,0) = k(\mu)k(-\mu)+2$. 
The mixed partial $\frac{\partial^2}{\partial\epsilon\partial\mu}$ 
vanishes at the origin since the $a,b$ coefficients are even functions
of $\mu$ for all $\epsilon$. Note that, because of the square root branch 
point, it is necessary to carry the expansion out to second order 
to get what amounts to a first order result - the local normal form 
is a cone. 

We present the calculation of 
$\frac{\partial\tr(\MM|{\mu=0})}{\partial\epsilon}$ first. 
The problem ${\bf L_-} \psi = 0$ has a periodic solution that we denote 
by $\psi_1(=\psi_{stand})$, and a linearly growing solution that 
we denote by $\psi_2$. 
We choose $\psi_{1,2}$ to form an orthogonal right-handed coordinate system:
\begin{eqnarray*}
\psi_1(0) = \cos(\theta) &~&~~ \psi_1^\prime(0) = \sin(\theta) \\
\psi_2(0) = -\sin(\theta) &~&~~ \psi_2^\prime(0) = \cos(\theta)
\end{eqnarray*}
Since $\mu=0$ is a band edge for the $\Lm$ operator we have the 
following expressions for the period map:
\begin{eqnarray*}
\psi_2(1)&=& \psi_2(0) + \sigma \psi_1(0) = \sigma \cos(\theta) - 
\sin(\theta)\\
\psi_2^\prime(1)&=& \sigma \sin(\theta) + \cos(\theta).
\end{eqnarray*}
Note that $\sigma$ is non-zero as long as the band edge is not a 
double point. 
It is convenient to define a second set of solutions $\phi_{1,2}$ which 
satisfy $\Lm \phi = 0$ along with the initial conditions 
$\phi_1(0)=1, \phi_1^\prime(0)=0$ and $\phi_2(0)=0, \phi_2^\prime(0)=1$.
These are obviously related to the $\psi_{1,2}$ by 
$\phi_1 = \cos(\theta) \psi_1 - 
\sin(\theta) \psi_2$ and  $\phi_2 = \sin(\theta) \psi_1 + \cos(\theta) \psi_2$.
These functions form a natural basis in which to do perturbation 
theory on the $\Lp$ operator. 
As is usual in Floquet theory we must construct a fundamental set of 
solutions $\tphi_1,\tphi_2$ to 
\begin{equation}
\Lp \tphi_{1,2} = \Lm \tphi_{1,2} + 2 \epsilon |\psi|^2 \tphi_{1,2} = 0
\end{equation}
that satisfy the boundary conditions 
\begin{eqnarray*}
\tphi_1(0) &=& 1 ~~~~ \tphi_1^\prime(0) = 0 \\
\tphi_2(0) &=& 0 ~~~~ \tphi_2^\prime(0) = 1 
\end{eqnarray*}
A straightforward perturbation calculation gives the following expressions 
for the fundamental set of solutions to $\Lp \tphi = 0:$ 
\begin{eqnarray*}
\tphi_1(x) &=& \phi_1 - 4 \epsilon \left(\psi_1 \int \psi_1^2 \psi_2 \phi_1 
- \psi_2 \int \psi_1^3\phi_1 \right) \\
\tphi_2(x) &=& \phi_2 - 4 \epsilon\left( \psi_1 \int \psi_1^2 
\psi_2\phi_2 - 
\psi_2 \int \psi_1^3\phi_2 \right) 
\end{eqnarray*}
Using the fact that $\psi_2(1)=\psi_2(0)+\sigma\psi_1(0)$ gives the 
following expression for the trace of the monodromy matrix to 
the leading order in $\epsilon:$ 
\begin{eqnarray*}
\tphi_1(1) &=& \phi_1(1) - 4 \epsilon \left( \cos^2(\theta) 
(<\psi_1^3 \psi_2> - 
\sigma <\psi_1^4>)\right.  \\
&+& \sin\theta\cos\theta(\sigma <\psi_1^3\psi_2> + <\psi_1^4> - 
<\psi_1^2\psi_2^2>) \\
&-& \left. \sin^2(\theta) <\psi_1^3\psi_2> \right) \\
\tphi_2^\prime(1) &=&\phi_2^\prime(1) - 
4\epsilon( -\cos^2(\theta) <\psi_1^3\psi_2> \\ 
&+& \sin(\theta)\cos(\theta)<\psi_1^2\psi_2^2> - \sigma<\psi_1^3\psi_2> - 
<\psi_1^4> )\cr 
&+& \sin^2\theta (<\psi_1^3 \psi_2> - \sigma <\psi_1^4>)
\end{eqnarray*}
where $<\!f\!> = \int_0^1 f(y) dy$. 
From this it follows that the monodromy matrix in leading order of 
$\epsilon$ is given by 
\begin{equation}
\tr(\MM|\mu=0)= \tphi_1(1) + \tphi_2^\prime(1) + \phi_1(1) + \phi_2^\prime(1)
= 2(\phi_1(1) + \phi_2^\prime(1)) + 4\epsilon \sigma <\psi_1^4> + O(\epsilon^2)
\end{equation}.

It is easy to calculate that $a(\mu,0) = -4 - k''(0)\mu^2 + o(\mu^2)$, and 
that $b(\mu,0) = 6 + (2 k''(0)-(k'(0))^2)\mu^2 + o(\mu^2)$. This, together 
with the vanishing of the mixed partial and the explicit formula 
(\ref{eqn:fact2}) gives the above result. 

{\bf Remark}  From the proof of the preceeding lemma, 
and the use of the time-invariant Hamiltonian energy functional
\begin{equation}
{\cal H}= \int \left(\frac{1}{2}|\psi_x|^2 \pm \frac{\epsilon}{2} |\psi|^4 - 
V(x)|\psi|^2\right) dx, 
\end{equation}
it follows that
\begin{equation}
\left.{\frac{\partial{\tr \MM}}{\partial \epsilon}}\right|_{(0,0)} = 
8 \sigma \left.{\frac{\partial\cal H}{\partial \epsilon}}\right|_{\epsilon=0}.
\end{equation}
Thus, the instability criterion can be expressed in terms of the signs of 
$\sigma$ and of the derivative of the energy functional. This is a 
feature that appears regularly in the study of stability 
of nonlinear waves: that the sign of the derivative of some conserved 
quantity is a proxy for the index of some linearized operator.  
For some results of a similar 
flavor in a variety of different contexts (both conservative
and dissipative) see \cite{Wein,GSS,Kap1,Zum,LP}.

\begin{thm}For $\epsilon$ small and positive (focusing NLS) and 
solutions bifurcating from a lower band edge that is not a 
double point, or for  $\epsilon$ small and 
negative and  solutions bifurcating from an upper band edge that is not 
a double point there exists a 
band of spectrum along the imaginary axis. 
\end{thm}

{\bf Proof:} The proof follows from the preceeding lemma and the same  
analyticity argument as the first theorem.
From the explicit factorization of the characteristic polynomial 
it follows that, for $\epsilon$ fixed, the Floquet discriminants 
$K_\pm(\mu)$ are analytic 
functions in the cut plane with branch points at the points where 
\begin{equation}
2 \tr(\MM^2) - \tr(\MM)^2 + 8 = 0
\end{equation}
From the preceeding Lemma it follows that at $\mu = 0$ this quantity 
reduces to 
\begin{equation}
\left(2 \tr(\MM^2) - \tr(\MM)^2 + 8\right)|_{\mu = 0} = ( \tr(\MM)|_{\mu=0} 
- 4)^2 = 16 \sigma^2 \epsilon^2 <\!\!\psi_1^4\!\!>^2 + o(\epsilon^2)
\end{equation}
and thus this quantity is nonzero for small but nonzero $\epsilon$: 
the fact that the band edge is not a double point guarantees that 
$\sigma \neq 0$. It follows that the Floquet 
discriminants $K_\pm$ are analytic in
a neighborhood of $\mu=0$ and satisfy
\begin{eqnarray*}
K_+(0) &=& 2 + 4 \epsilon \sigma <\psi_1^4> + O(\epsilon^2)\\
K_-(0) &=& 2.
\end{eqnarray*}
As in the first theorem the fact that $K_+(\mu)$ is an even function 
implies that $\mu = 0$ is a critical point, and $K_+$ is real on a segment 
of the imaginary axis containing the origin. 
Recall that for odd numbered bands the periodic eigenvalue is the 
lower band edge, and $\sigma$ is positive, and for even numbered
bands the periodic eigenvalue is the upper band edge, and $\sigma$ 
is negative. 
For $\epsilon<0$ (defocusing) and odd numbered bands, 
or for $\epsilon>0$ and even numbered bands the perturbation result 
implies that $K_+(0) \in (-2,2)$, so there exists an interval on the 
imaginary axis on which $K_+(\mu)$ is real and $\in (-2,2)$. 
The analogous calculation for solutions bifurcating from the 
anti-periodic eigenvalues shows that for $\epsilon<0$ (defocusing) and
even numbered bands, for $\epsilon>0$ and odd numbered bands $K_+(0) 
\in (-2,2)$, and there exists a spine of spectrum along the imaginary axis. 
Thus, for $\epsilon$ small and 
defocusing nonlinearity standing wave solutions which bifurcate 
from the lower band edge are modulationally unstable, while for 
focusing nonlinearity  standing wave solutions which bifurcate 
from the upper band edge are modulationally unstable. 
Said more 
simply the solutions which bifurcate from the lower band
edges are modulationally unstable in the defocusing case, 
and the solutions which bifurcate from the upper band edges 
are modulationally unstable in the focusing case. 

Note that if $K_\pm^{\prime\prime}(0)=0$ there exist additional 
arcs of spectrum emerging from the origin into the complex plane. 
For $\epsilon$ sufficiently small the local normal form calculation 
guarantees that the second derivative is 
non-vanishing, and this does not occur. It is a possibility 
for larger $\epsilon$, however.

\section{Sideband Instabilities}

In this section we consider the possibility of sideband 
type instabilities. This case is somewhat more difficult 
than the case of modulational instabilities, since 
there are fewer symmetries, and our results are 
somewhat less detailed. In addition to the analyticity 
arguments of the previous section a major tool will be 
the Krein signature. 

First we note the following lemma 
\begin{lem}
The spectrum of the stability problem consists of 
a union of continuous curves in the complex plane. 
The possible endpoints of the curves are 
band edges $K_\pm(\mu) = \pm 2$, branch points $a^2(\mu) - 4 b(\mu) + 8 =0$
(or $K_+(\mu) = K_-(\mu)$) or critical points $K_\pm^\prime(\mu)=0$.
\end{lem}

{\em Proof:} As before $K_\pm(\mu)$ are analytic away from 
the (isolated) points where   $a^2(\mu) - 4 b(\mu) + 8 =0$. 
Suppose $\mu$ is a point in the spectrum that is not a 
branch point, band edge or critical point. We assume for the 
sake of argument that $K_+(\mu) \in (-2,2)$. Since $K_+$ is 
analytic in a neighborhood of $\mu$ and $\mu$ is not a 
critical point of $K_+$ an appeal to the implicit function theorem 
shows the existence of a unique curve through $\mu$ along 
which ${\cal I}m(K_+)=0.$ Since $\mu$ is not a band edge 
continuity implies that ${\cal R}e(K_+)\in(-2,2)$ in some neighborhood
of $\mu$. 

We now consider the possibility of sideband instabilities arising 
from points where $K_+(\mu) = K_-(\mu)$. First note that in the 
small amplitude limit, when $\epsilon=0$, we have $a^2 - 4 b + 8 = 
(k(\mu)-k(-\mu))^2$ so the branch points are degenerate. Under perturbation 
this degeneracy is expected to break, and generically it can do so in one 
of two ways, which are generally referred to as the avoided collision
and the open gap. These possibilities are illustrated in figure 
\ref{fig:posbif}.
It is rather tedious to compute the normal form for this case,
since one lacks the even symmetry, but consideration of the 
Krein-Gelfand-Lidskii theorem can reduce the possibilities. In the case 
of a collision of two eigenvalues with the same Krein sign the open 
gap is forbidden, since that corresponds to eigenvalues leaving the 
unit circle. In the case of two eigenvalues with the opposite Krein 
sign one expects that under perturbation the eigenvalues should 
leave the unit circle. We formalize this observation as a theorem. 

\begin{thm}
In the neighborhood of an accidental degeneracy ($\mu = \mu^*, \epsilon=0$)
the Floquet discriminants have the following normal form
\begin{eqnarray*}
&K_\pm(\mu) = k(\mu^*) +  (k'(\mu^*) - k'(-\mu^*))(\mu-\mu^*) + 
\alpha \epsilon + E_1 \pm & \\
&\sqrt{\left(k'(\mu^*)+k'(-\mu^*)\right)^2 (\mu-\mu^*)^2  + \gamma \epsilon^2 
+ E_2}& \\
& E_1 = O(\epsilon^2, (\mu-\mu^*)^2,\epsilon(\mu-\mu^*))& \\
& E_1 = O(\epsilon^3, (\mu-\mu^*)^3,\epsilon(\mu-\mu^*)^2, 
\epsilon^2(\mu-\mu^*))&
\end{eqnarray*}
Further, if $\sgn(k'(\mu^*)) = \sgn(k'(-\mu^*))$ the quadratic form 
$(k'(\mu^*)+k'(-\mu^*))^2 (\mu-\mu^*)^2 + \beta \epsilon (\mu- \mu^*) + 
\gamma \epsilon^2$ is nonnegative. 
\end{thm}  
A straightforward, though slightly tedious, perturbation argument shows that 
$\frac{\partial}{\partial \epsilon} (a^2 - 4b + 8)|_{\epsilon =0}=0$, so 
that this quantity is locally quadratic, as well as having a vanishing 
mixed partial. The Krein signs of the unperturbed 
eigenvalues are the same as the signs of $k'(\mu)$ and  $k'(-\mu)$
respectively. In the case where these signs are the same the 
Krein-Gelfand-Lidskii theorem guarantees that, under small perturbations, 
the eigenvalues of $\MM$ remain on the unit circle. It is clear that
a necessary condition for the eigenvalues to remain on the unit circle is 
non-negativity of the quadratic form, thus the sign condition guarantees 
non-negativity of this form. Note that this theorem 
does not preclude the following possibilities: in the case  of like Krein 
sign collisions it is possible that the quadratic form has a zero 
eigenvalue, or that the intersection persists under perturbation. 
In the case of collisions of opposite Krein sign a naive application 
of the Krein-Gelfand-Lidskii theorem gives no information. We conjecture 
that, in this case, the signature of the quadratic form is equal to the
signature of the subspace. This would imply that that intersections of 
opposite Krein signs always open to a gap, while intersections of the same 
Krein sign always open to an avoided collision.
Very preliminary numerical evidence has 
supports this conjecture, but unfortunately it appears that the only way 
to verify this analytically is via a very tedious perturbation argument. 
 
An obvious corollary of this is the following:

\begin{corr}
If $\sgn(k'(\mu^*))=\sgn(k'(-\mu^*))$ and $\gamma > 0$,
or if $\sgn(k'(\mu^*))=-\sgn(k'(-\mu^*))$ and 
$\gamma<0$ then there exists a band of spectrum 
off of the real axis.

\end{corr}

From the local normal form it is an easy calculation that, in either case, 
for sufficiently small $\epsilon$ the Floquet discriminants $K_\pm$ must 
have a critical point in the neighborhood of $\mu^*$, and that 
$a^2 - 4 b + 8 \neq 0$ at this critical point. From the analyticity 
arguments of the previous section this guarantees that $K_\pm$ are 
real and $\in (-2,2)$ in some neighborhood of the critical point. 
Again one expects that in the case of a like Krein sign collision 
one should generically have $\gamma>0$, leading to the avoided 
collision and a loop of spectrum opening into the complex plane, but 
there seems to way to show this in any particular case without
actually calculating $\gamma$.

\section{Explicit Examples and Numerics}

In this section we present some examples. We will primarily be working with 
known exact elliptic function solutions. We consider the 
nonlinear Schr\"odinger equation
\begin{equation}
i\psi_t = -\frac{1}{2} \psi_{xx} + V_0 \sn^2(x,k) \psi \pm |\psi|^2 \psi\end{equation}
This equation has a one-parameter family of exact solutions 
given by 
\begin{eqnarray}
\psi(x,t) &=& r(x) e^{-i\omega t + i \theta(x)} \\
r^2(x) &=&  A \sn^2(x,k) + B \\
\theta(x) &=& c \int\frac{dx'}{r^2(x')} \\
A &=& -(V_0 + k^2) \\
c^2 &=& B\left( \frac{B}{V_0 + k^2 -1}\right)\left(V_0 + k^2 - B k^2\right)
\end{eqnarray}
Where $B \in \left(-\infty,-k^2\right) \cup 
\left(\frac{V_0 + k^2}{k^2}, V_0 + k^2\right)$
for the focusing sign  and $B \in  \left(-(V_0 + k^2),-\frac{V_0 + k^2}{k^2} 
\right)\cup \left(-k^2,\infty\right)$ for the defocusing sign. These 
solutions 
represent nonlinear stationary states which bifurcate from the linear 
Bloch states. We are primarily interested in the solutions which bifurcate 
from the band edges. These correspond to the boundaries of the 
above regions of validity. In the focusing case we have 
\begin{eqnarray}
r_0(x) &=& \frac{\sqrt{V_0 + k^2}}{k^2} \dn(x,k) ~~~ \omega = -1 - 
\frac{V_0}{k^2} + 
\frac{k^2}{2} ~~~~ V_0 + k^2 > 0\\
r_1(x) &=& \sqrt{V_0 + k^2}\cn(x,k) ~~~ \omega = \frac{1}{2} - V_0 - k^2 ~~~ 
v_0 + k^2 > 0\\
 r_2(x) &=& \sqrt{-(V_0 + k^2)}\sn(x,k) ~~~ \omega = \frac{1+k^2}{2} ~~~ V_0 + 
k^2 < 0,
\end{eqnarray}
while in the defocusing case we have the analogous solutions
\begin{eqnarray}
r_0(x) &=& \frac{\sqrt{V_0 + k^2}}{k^2} \dn(x,k) ~~~ \omega = -1 - \frac{V_0}
{k^2} + 
\frac{k^2}{2} ~~~~ V_0 + k^2 < 0\\
r_1(x) &=& \sqrt{V_0 + k^2}\cn(x,k) ~~~ \omega = \frac{1}{2} - V_0 - k^2 ~~~ 
v_0 + k^2 < 0\\
 r_2(x) &=& \sqrt{-(V_0 + k^2)}\sn(x,k) ~~~ \omega = \frac{1+k^2}{2} ~~~ 
V_0 + k^2 > 0.
\end{eqnarray}
The solutions all bifurcate from the the linear Bloch states at 
$V_0 + k^2 = 0$, with the $\dn,\cn$ solutions existing 
on one side of the bifurcation and the $\sn$ solution on 
the other side. 
It is easy to check that the spectrum of the 
${\bf L_-}$ operator in each of these cases is the 1-gap Lam\'e operator
plus some constant which differs in each case. The ground state of 
${\bf L_-}$ is $\dn(x,k)$, while the next two antiperiodic eigenfunctions
are given by $\cn(x,k), \sn(x,k)$.
In what follows the parameter $\epsilon = V_0 + k^2$. There is 
an obvious rescaling which transforms the above form of the 
NLS equation to the form considered earlier.  

\subsection{Focusing Case:}

It follows from Theorem (2) that for $V_0 + k^2 = \epsilon$ suffiently
small that the $\dn$ and $\sn$ solutions, as lower band edges, 
are modulationally unstable. 
Moreover, from Theorem (1) we are guaranteed a modulational instability 
as long as the ${\bf L_+}$ operator is in the interior of a band 
at $\mu=0.$ Thus in each case we have an open interval $(0,\epsilon^*)$ 
in which we are guaranteed a modulational instability, where $\epsilon^*$
is the smallest positive value of $\epsilon$ such that $\mu=0$ is a band 
edge of the ${\bf L_+}$ operator. We will focus on the $\sn$ solution, 
since the instability of the $\dn$ solution follows from more 
elementary arguments\cite{}. We have found numerically that
$\mu = 0$ is in the interior of a band for $\epsilon 
\in (-1.33,0)\cup(-3.0,-1.56)
\cup(-6.7,-5.6)\ldots$, which shows instability for $\epsilon$ 
in these intervals. The numerical evidence further shows that in this 
particular case the second derivative of 
the branch of the Floquet discriminant passing through $-2$ is always 
negative, which implies that these solutions are always unstable.
We do not currently have a proof of this. It is also unclear whether 
this feature is special to the elliptic function solutions, or 
if it holds in greater generality. It is interesting to note that 
the $\sn(x,k)$ solution also has a side-band type instability which 
appears at arbitrarily small positive amplitude. This is illustrated 
in \ref{fig:finiteamp}. 
When $\epsilon = 0$ the Floquet discriminants of the 
unperturbed problem $k(\pm \mu)$ intersect near $\mu \approx .45.$ 
This is a collision of opposite Krein sign, since the discriminants 
have the same sign. This intersection opens into a gap under 
perturbation. 
 
\begin{figure}
\begin{center}
\begin{tabular}{cc}
\psfig{file=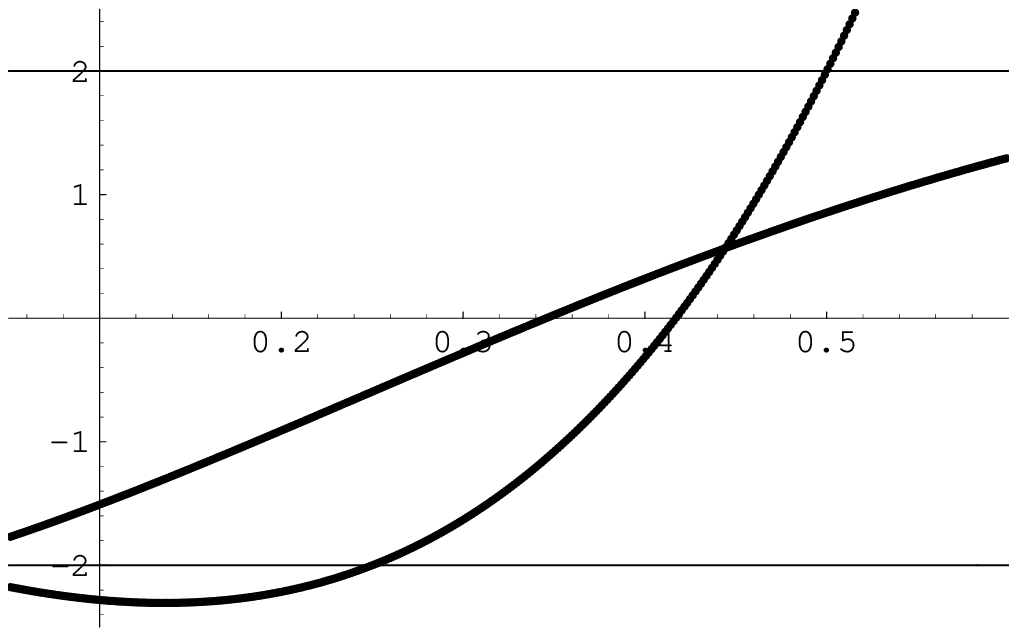,height=1.3in} & 
\psfig{file=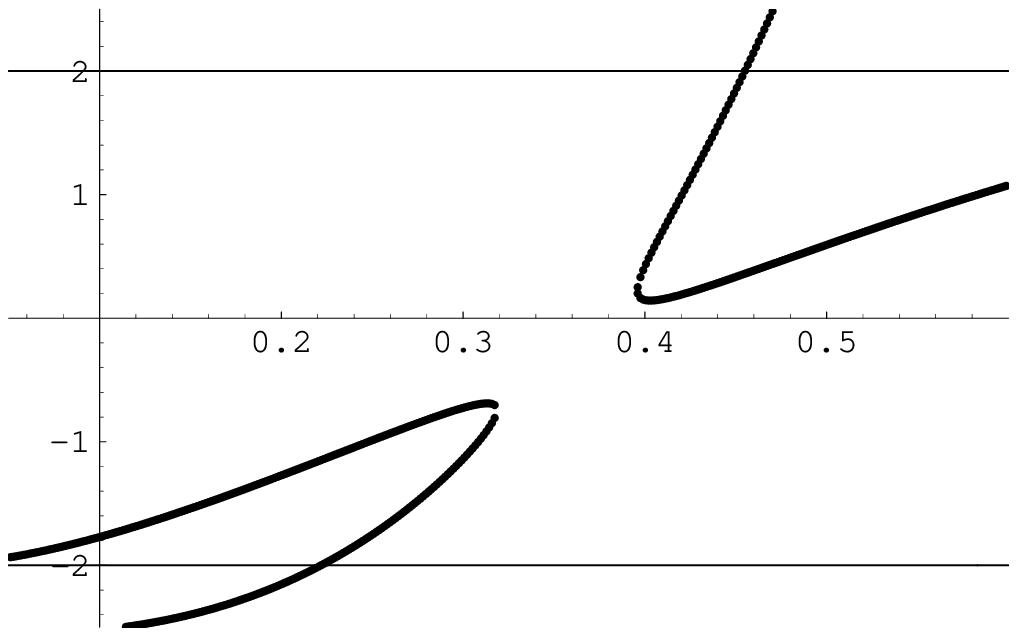,height=1.3in}\end{tabular}
\end{center}
\caption{ The Floquet Discriminants for the $\sn(x,k)$ solution in the 
focusing case with 
$\epsilon=0,.1$}
\label{fig:finiteamp}
\end{figure} 

The stability of the $\cn$ type solutions is extremely interesting. 
For small $\epsilon$ the perturbation result guarantees that 
$\mu=0$ is in a gap for the ${\bf L_+}$ operator, and numerically 
we find no modulational instability for small $\epsilon$. 
As before we expect that for some critical value of $\epsilon$ 
$\mu = 0$ will enter a band of ${\bf L_+}$ and an instability will 
develop. In this case the critical value of $\epsilon$ can 
actually be computed explicitly. When $V_0=0
(\epsilon = k^2)$ we find that ${\bf L_+}$ is the 2-gap Lam\'e operator, 
with $\mu=0$ is a band edge. It is also straightforward to show 
using the Sturm oscillation theorem and monotonicity of the 
band edges in $\epsilon$ that 
this is the smallest value of $\epsilon$ for which ${\bf L_+}$ 
has a band edge at $\mu=0$. Thus we are guaranteed a modulational instability
for some interval of beginning at $V_0 = 0 (\epsilon = k^2)$. 
It is worth remarking that in this critical case (the integrable 
one) all of the information about the spectrum of the linearization,
as well as a great deal more, can be obtained by the methods of algebraic
geometry. See, in particular, Chapter 4 of the text by Belokolos, Bobenko, 
Enol'skii, Its and Matveev\cite{BBEIM}. The birth of a modulational 
instability at $V_0=0$ supports a intuition based on physical reasoning
that was put forth in \cite{BCCDKP2} which suggested  
that such solutions should go unstable at $V_0=0$.

Interestingly  it appears that there exists a finite amplitude 
side-band type instability that sets in before the modulational 
instability. This is illustrated in figure \ref{fig:finiteamp}, which 
shows the Floquet discriminants for the linearized operator of the 
focusing NLS about a $\cn$ type solution with modulus $k^2 = 1/2$
and a sequence of different values of $\epsilon$.
 Near $V_0 \approx -.23$ the Floquet
discriminants $K_\pm$ cross, causing a gap to open and a 
loop of spectrum to emerge into the complex plane. This instability 
disproves a guess made in \cite{BCCDKP2}, based on numerical experiments, 
that the $\cn$ type solutions should be stable for $V_0<0$.
As $V_0$ increases the neighborhood about the origin in which the 
Floquet discriminants are real shrinks, until at $V_0=0$ the 
discriminants are only real at the origin. For positive $V_0$ the 
discriminants again become real in a neighborhood of the origin.
 
\section{Conclusions}

We have established a sufficient condition for the modulational 
instability of a periodic standing wave solution to the NLS 
equation, which can be easily checked. In the case of weak 
nonlinearity this reduces to a physically reasonable criterion 
based on the effective mass of a particle in the periodic potential. 
We have also made some preliminary  progress into studying the 
birth of side-band type instabilities. In the case of like 
Krein-sign collisions the gap-opening bifurcation is forbidden.
Very preliminary numerical experiments suggest that, in the 
case of opposite Krein-sign collisions the ``avoided collision'' 
bifurcation is forbidden, but we do not currently have a proof 
of this. 

{\bf Acknowledgements:} JCB is partially supported by NSF-DMS 0203938
and an NSF FRG grant.

\begin{figure}
\begin{center}
\begin{tabular}{cc}
\psfig{file=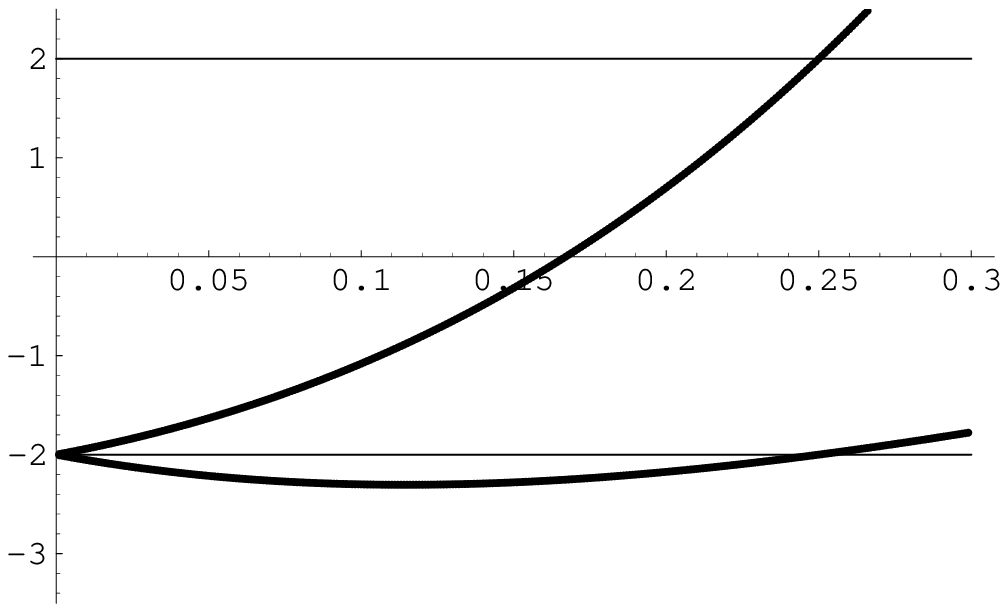,height=1.1in} & 
\psfig{file=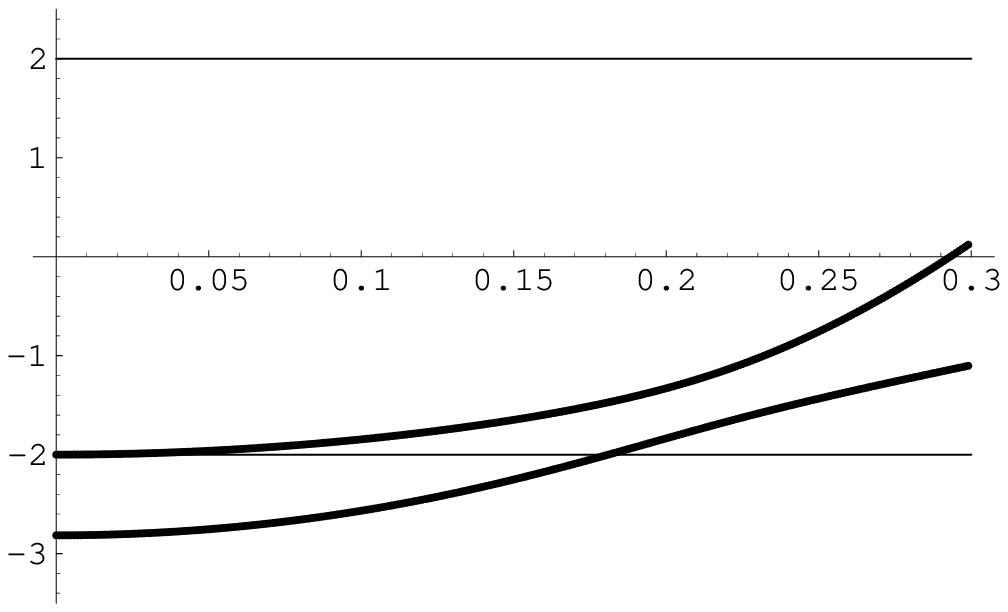,height=1.1in} \\
\psfig{file=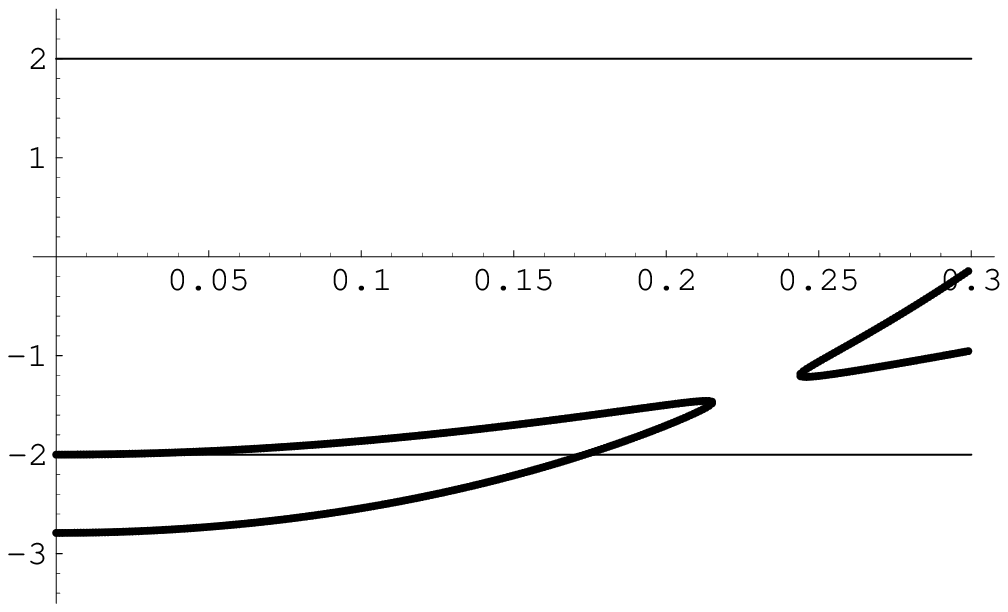,height=1.1in} & 
\psfig{file=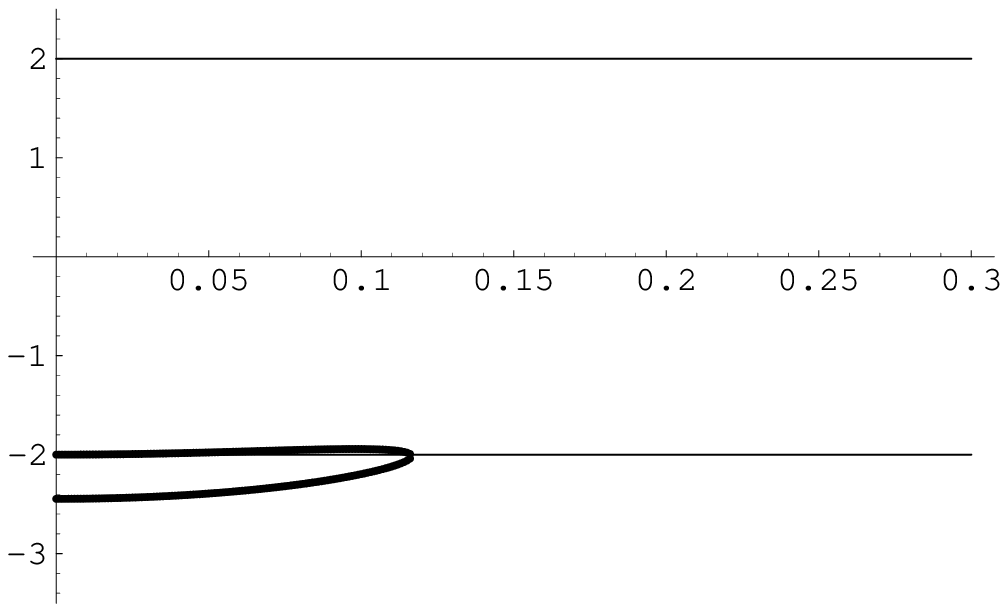,height=1.1in} 
\end{tabular}
\end{center}
\caption{ The Floquet discriminants for the focusing $\cn(x,k)$ 
solution for $V_0 = -1/2, 
-1/4, -.228, -.1$. }

\label{fig:finiteamp}
\end{figure}

\begin{figure}
\begin{center}
\begin{tabular}{cc}
\psfig{file=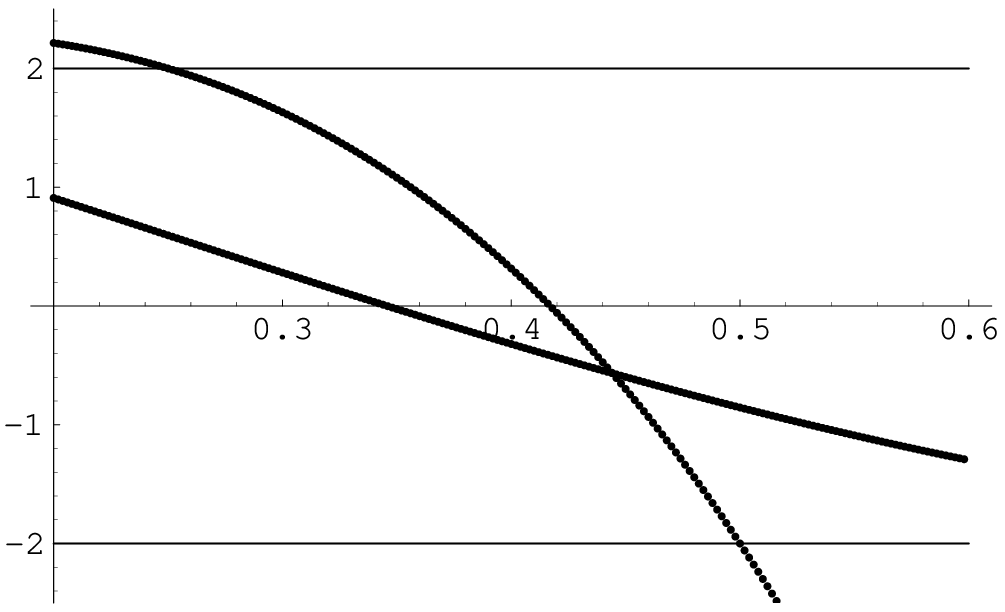,height=1.35in}& \\
\psfig{file=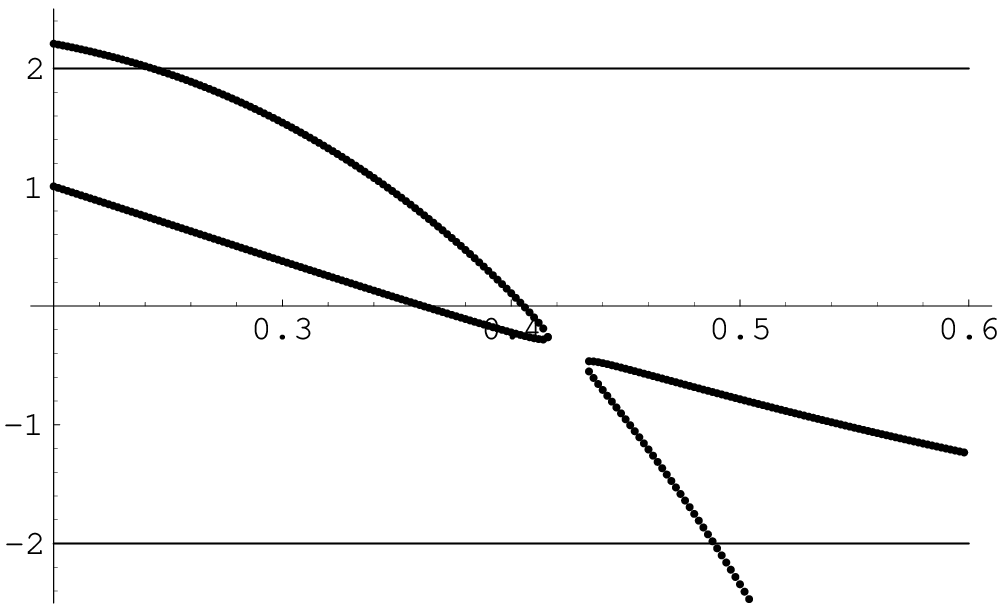,height=1.35in}&  \\
\psfig{file=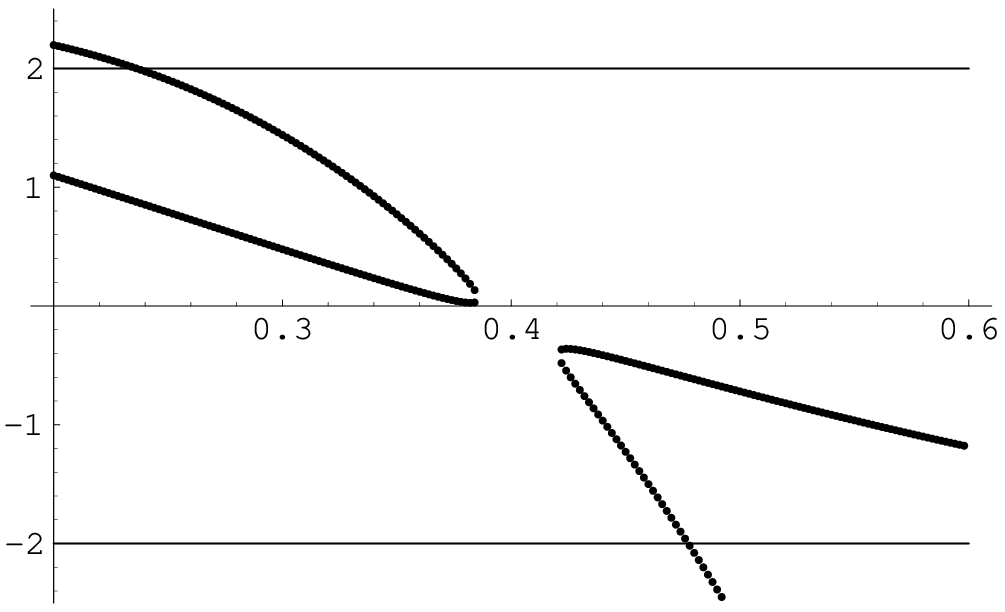,height=1.35in}& \\
\psfig{file=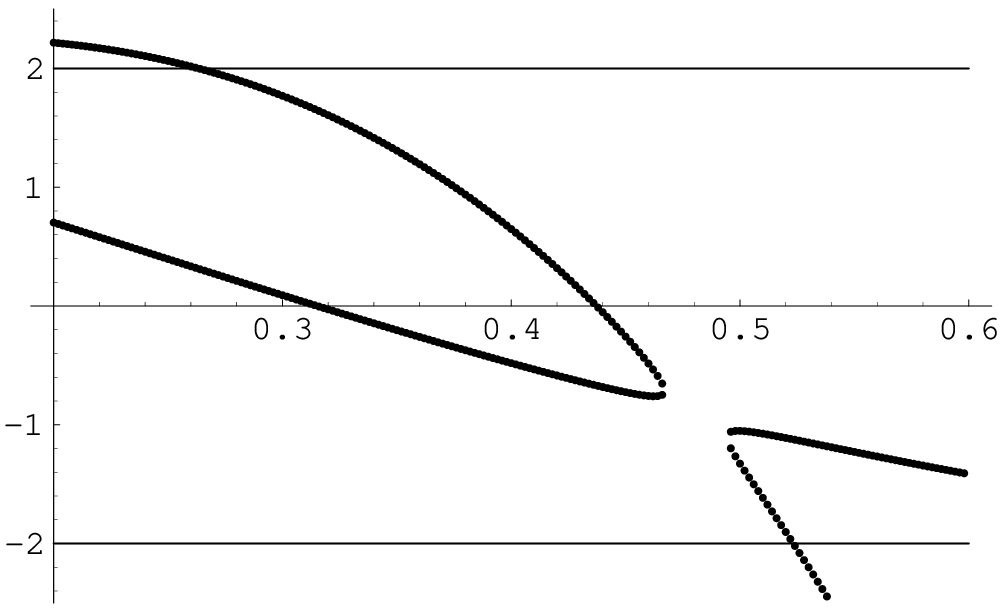,height=1.35in}&  

\end{tabular}
\end{center}
\caption{A perturbed Floquet spectrum where the unperturbed problem has 
degenerate eigenvalues of opposite Krein signs. The graphs show the 
Floquet discriminants along the real $mu$ axis for $\epsilon=0,0.025,0.05,
-0.05$. 
Note the opening of a small gap in the neighborhood of the intersection.}

\label{fig:badkrein}
\end{figure}

\begin{figure}
\begin{center}
\begin{tabular}{cc}
\psfig{file=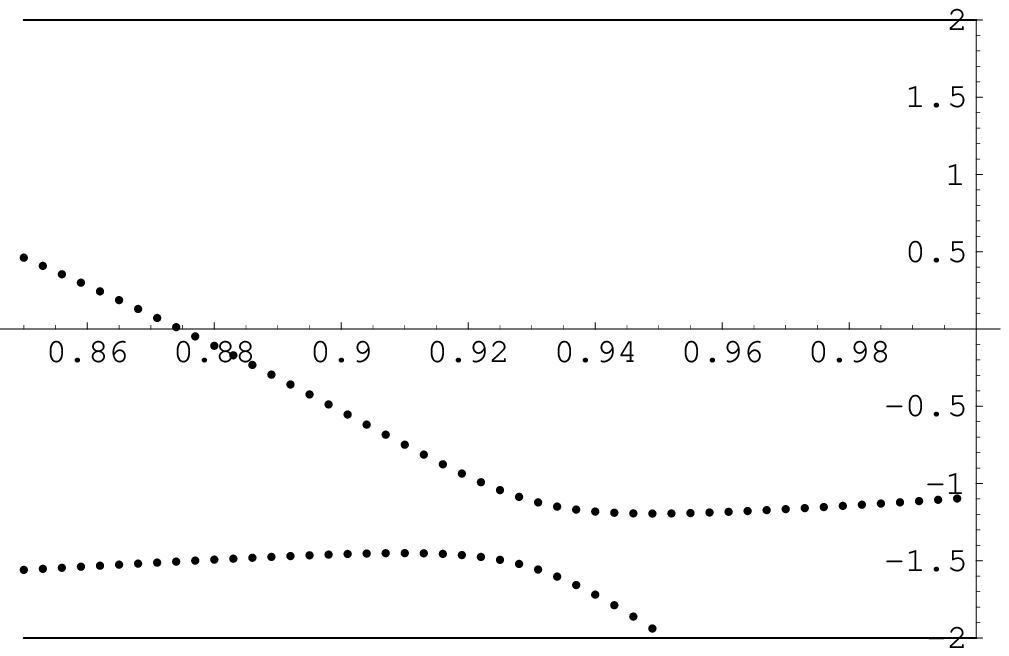,height=1.75in}&\psfig{file=
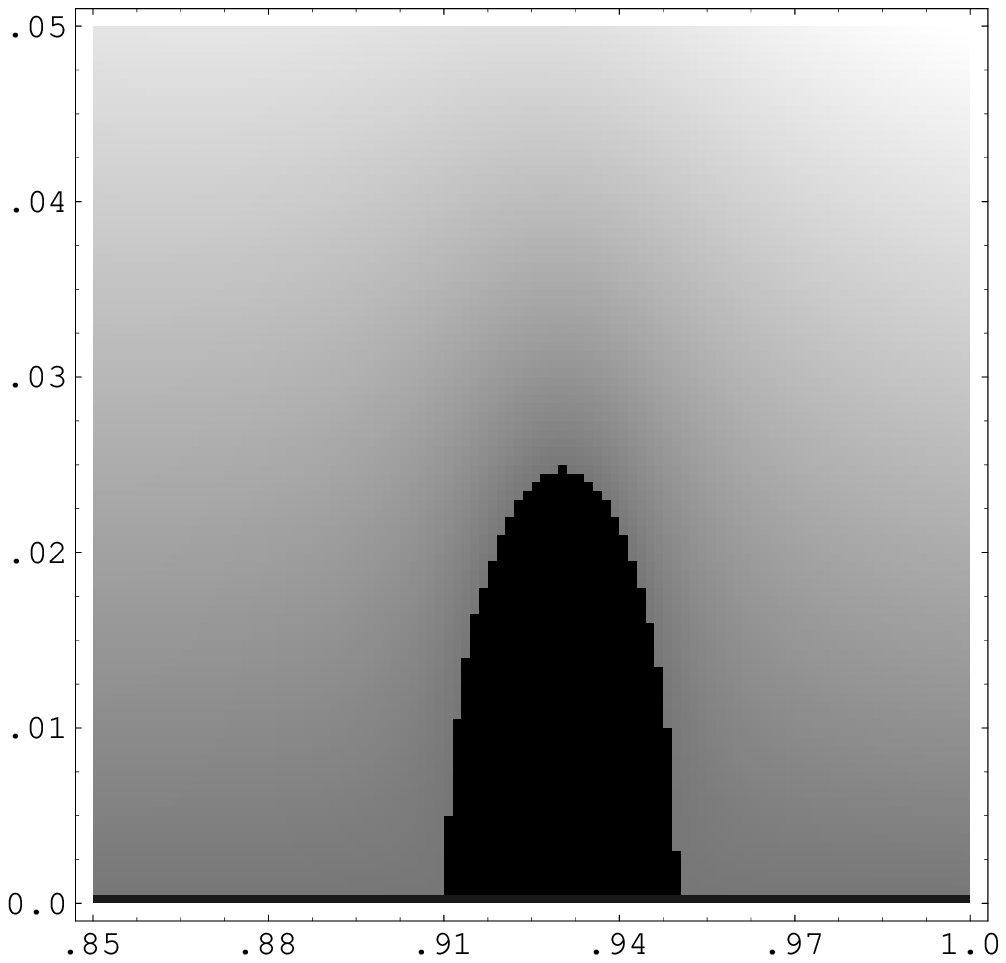,height=1.75in} 
\end{tabular}
\end{center}
\caption{A perturbed Floquet spectrum where the unperturbed problem has 
degenerate eigenvalues of the same Krein sign for a particular value of 
$\mu$ along the real axis. The first figure shows the Floquet discriminants 
along the real $\mu$ axis (note the avoided collision) and the second 
the imaginary part of the floquet discriminant in the complex plane. 
The imaginary part of the discriminant is negative inside the eye and 
positive outside.}

\label{fig:goodkrein}
\end{figure}

\begin{figure}
\begin{center}
\begin{tabular}{c}
\psfig{file=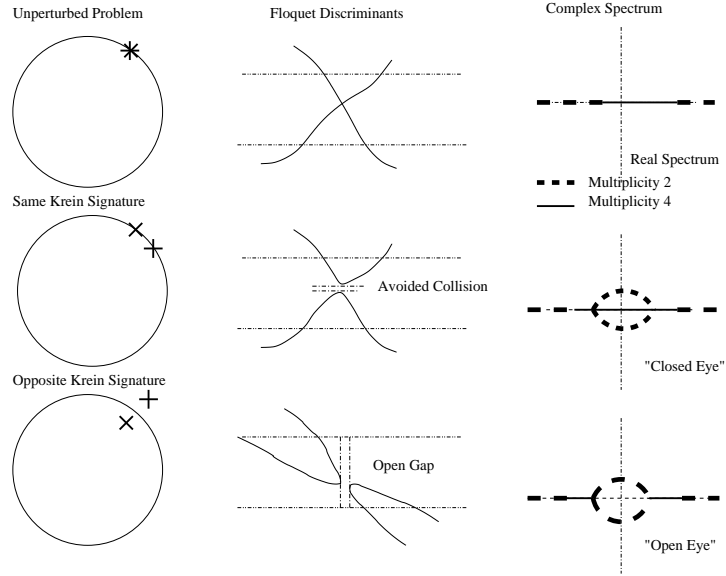,height=3.0in}
\end{tabular}
\end{center}
\caption{The possible bifurcations of degenerate eigenvalues of a symplectic
matrix under perturbation, and the local behavior of the Floquet 
discriminants.}

\label{fig:posbif}
\end{figure}

\bibliography{pmibiblio}
\end{document}